\documentclass[11pt,a4paper]{article}
\usepackage[margin=1in]{geometry}
\usepackage{iftex}
\ifPDFTeX
  \usepackage[T1]{fontenc}
  \usepackage[utf8]{inputenc}
  \usepackage{lmodern}
\else
  \usepackage{fontspec}
\fi
\usepackage{amsmath,amssymb}
\usepackage{booktabs,longtable,array}
\newcolumntype{P}[1]{>{\raggedright\arraybackslash}p{#1}}
\usepackage{url}
\usepackage{graphicx}
\usepackage[round,authoryear]{natbib}
\usepackage{microtype}
\usepackage[hidelinks]{hyperref}
\usepackage{xcolor}
\title{Trainable Quantum-Native Language Models:\\Dissipative Memory, Capacity Limits, and Hardware Validation}
\author{Deniz Yunus G\"o\u{g}\"u\c{s}}
\date{Preprint, October 2026}

\begin{document}
\maketitle

\begin{abstract}
We study byte-level autoregressive language models whose hidden state is a quantum state on a small register, whose state transition is a learned, token-conditioned quantum channel, and whose next-byte distribution is read out directly as Born-rule measurement probabilities, with no classical vocabulary projection or softmax. Within this fixed-width recurrent architecture, a purely unitary memory cannot contract or reset stored distinctions, and in our experiments it loses to its own retrained one-token control in every variant tested; interposing a learned dissipation channel, and then making the dissipation token-conditioned (an input-conditioned CPTP map playing the role of a recurrent gate), yields models whose memory demonstrably carries information useful beyond one token (+0.20 $\pm$ 0.03 nats over one-token controls of the same parameterization and parameter count, retrained from scratch, at context 12; +0.29 to +0.33 at longer contexts), while never surpassing a retrained two-token window. On a key--value recall benchmark the largest solvable task shifts upward with register width over 4--7 qubits (one stored pair at 4--5 qubits, two at 6--7, three unsolved), and the persistent two-token horizon on text is consistent with an information-capacity bottleneck of the scale set by the Holevo bound; we present this as an empirical threshold shift over the tested range, not as a capacity law. As a secondary result, one-qubit models trained on a perturbed-coin process carry less stationary memory entropy than any classical predictor found by a Lagrangian attack at matched prediction error (emergently in 9/9 seeds, median margin 0.04 nats; 0.44--0.66 nats with an annealed entropy penalty), a memory-entropy separation claimed strictly within that scope. Compiled to hardware-shaped circuits, the model's computational-basis memory-register distribution was validated on a 156-qubit IBM Heron processor at every prefix length up to its full 12-token context (12/12 circuits closer to the simulator's predicted distribution than that prediction is to uniform), its next-bit probabilities passed pre-registered accuracy bars at 4-token prefixes, and bytes were generated autoregressively with every bit sampled from physically measured probabilities. No computational quantum advantage is claimed: all models studied are classically simulable at these widths, and none approaches classical n-gram quality. What the work establishes is architectural and physical - which memory dynamics a trainable quantum-native recurrent model requires, how sharply small quantum memories are capacity-limited, and that the resulting model runs end-to-end on present hardware.

\end{abstract}

\section{Introduction}\label{sec:intro}
Can a quantum system serve as the \emph{native substrate} of a language model - state, dynamics, and sampling all quantum, with classical computation confined to optimization and control? The question is usually approached theoretically. We approach it experimentally, within a single architecture family, under a pre-registered, falsification-first protocol: every hypothesis was registered with quantitative bars before its runs, every claim was adjudicated against retrained parameter-matched controls and recomputed baselines, and every negative result is reported.

\textbf{Scope and the central statement.} The paper's main line is a sequence: closed (unitary) quantum memory $\to$ its inability to learn useful forgetting $\to$ learned dissipation $\to$ token-conditioned dissipative gating $\to$ measured capacity limits $\to$ compilation and validation on real hardware. The statement it supports is this: \emph{trainable quantum-native recurrent language models of this kind require controlled non-unitary memory dynamics; those dynamics produce useful recurrent memory but remain strongly capacity-limited at small qubit counts; and the resulting model can nevertheless be compiled and run end-to-end on present quantum hardware, with its measured statistics validated against simulation.} A memory-entropy result on a synthetic process is reported as a secondary contribution with a strictly stated scope.

\textbf{What "quantum-native" means here.} We call a model quantum-native when (i) its recurrent hidden state is a quantum state (a density matrix $\rho_t$ on n memory qubits), (ii) each state transition is a quantum channel (a completely positive trace-preserving map), (iii) the output distribution over tokens is obtained directly from measurement probabilities, $p(y) = \mathrm{Tr}(M_y\rho_t)$ for a POVM $\{M_y\}$, (iv) sampled outputs are fed back autoregressively as inputs to the quantum transition, and (v) there is no learned classical vocabulary projection producing logits. Classical optimization of the parameters, classical control flow, circuit compilation, shot counting, and sampling orchestration do not affect this definition; they are present in every quantum algorithm. The definition separates this work from hybrid designs in which a quantum circuit acts as a feature extractor whose measured expectation values feed a classical output layer (\S2).

\textbf{No computational advantage.} We state up front, and repeat in the limitations, that no computational quantum advantage is claimed. Classically, the models are input-switched Kraus-map state-space models with a quadratic readout; at 4--8 qubits they are exactly simulable, and all training in this paper was done in simulation. None of the models approaches the quality of a classical 4-gram model on the same corpus.

\textbf{Contributions.}

\begin{itemize}
  \item \textbf{A. Architecture.} We give a precise statement of why closed unitary recurrence cannot implement the many-to-one contraction that learned forgetting requires (\S4), show experimentally that unitary memory loses to retrained one-token controls in five variants at two widths, and show that a learned dissipation channel (UMW-D) and then a token-conditioned dissipative channel (UMW-G) restore and then more than double the beyond-one-token value of memory, with controls that are parameter-count-matched by construction (\S5).
  \item \textbf{B. Capacity.} On a key--value recall ladder over 4--7 qubits, the largest solvable task shifts from one stored pair to two as width grows, while three pairs (10 bits of value content, 19 bits of configuration entropy) remain unsolved at every width; we define the quantities involved and relate the models' persistent two-token horizon on text to accessible-information bounds, using the language of consistency rather than causation (\S6).
  \item \textbf{C. Hardware.} Compiled to line-entangler circuits, the model's computational-basis memory-register statistics, next-bit probabilities, and autoregressive generation were validated on an IBM Heron processor in three clearly separated levels, with local unitary-equivalence and simulator checks before every submission and every failure recorded (\S8).
  \item \textbf{Secondary.} A stationary memory-entropy separation from attacked classical predictors at matched error on a process whose classical statistical complexity is known exactly (\S7); a purification-factor training representation, exact at full rank, that enables controlled low-rank training at 7--8 qubits, together with scaling observations on compiled circuits (\S9, Appendix~\ref{app:a}).
\end{itemize}

\section{Related work}\label{sec:related}
\textbf{Quantum simulation of stochastic processes and quantum statistical complexity.} \citet{gu2012quantum} showed that a quantum simulator of a classical stochastic process can require less memory entropy than the statistical complexity $C_\mu$, which lower-bounds every exact classical predictor; subsequent work characterized the quantum statistical complexity and its optima \citep{mahoney2016occam,binder2018practical,loomis2019strong,elliott2020extreme}. These constructions are analytic. Our \S7 asks whether gradient-trained models land in that regime, using the perturbed coin whose $C_\mu = \ln 2$ is exact and whose quantum optimum $C_q$ is known in closed form.

\textbf{Quantum recurrent neural networks.} Recurrent quantum models have been proposed in several forms: parameterized circuits with a reset-and-reload cell \citep{bausch2020recurrent}, quantum long short-term memory cells in which variational circuits replace the gates of a classical LSTM \citep{chen2022quantum}, and quantum recurrent networks for sequential learning \citep{li2023quantum}. To our reading, these designs carry a \emph{classical} recurrent state between steps - measured expectation values that are re-encoded into the next circuit - and produce outputs through a classical layer. Closest to us in spirit is the fully quantum-native recurrent network of \citet{huang2026fully}, which preserves a quantum memory state across a sequence under coherent unitary recurrence and reports NISQ hardware experiments on time-series forecasting; it produces scalar forecast outputs rather than a distribution over a token vocabulary, and its recurrence is unitary - the very regime our \S4 finds unable to learn forgetting for language modeling. Persistent quantum recurrent state on hardware is therefore not itself new. Ours keeps the quantum state as the persistent hidden state, never measured between tokens, evolves it by \emph{learned dissipative, token-conditioned} channels, and obtains the token distribution directly from a POVM on that state. Quantum reservoir computing \citep{fujii2017harnessing,martinez2021dynamical} is a closer dynamical cousin: it exploits dissipative quantum dynamics precisely because fading memory is needed for sequential processing, but its readout is a trained classical linear map on measured observables and its dynamics are not trained.

\textbf{Quantum natural language processing.} The compositional (DisCoCat) program maps grammatical structure to tensor contractions and then to circuits \citep{coecke2010mathematical}, with hardware experiments on sentence classification and question answering \citep{meichanetzidis2023grammar,lorenz2023qnlp,kartsaklis2021lambeq}. These are classification or similarity tasks over fixed-length inputs rather than autoregressive sequence modeling.

\textbf{Quantum generative modeling and language generation.} Quantum circuit Born machines sample from a measured output distribution \citep{benedetti2019generative,liu2018differentiable,coyle2020born}, and hardware demonstrations of quantum generative models exist \citep{zoufal2019quantum,rudolph2022generation}. Born machines share with our readout the principle that probabilities are measurement probabilities, but they model a single fixed-length distribution rather than a recurrent autoregressive process with a persistent hidden state. Quantum natural language \emph{generation} on near-term devices has been demonstrated by sampling sentences from circuits that encode grammatical and semantic constraints \citep{karamlou2022quantum}; that approach does not train a recurrent model on a corpus and does not carry a quantum hidden state across generated tokens. \citet{balauca2025practical} report practical hybrid quantum language models - QRNN- and QCNN-style circuits run on real IBM hardware - with generative language modeling; their output path is hybrid: quantum recurrent processing $\to$ measured observables $\to$ a classical vocabulary projection producing logits. Quantum transformer proposals \citep{khatri2024quixer} target the attention computation and likewise retain a classical output head. The architectural distinction we draw throughout is between \emph{a quantum feature extractor feeding classical logits} and \emph{direct quantum-state-to-token-probability generation with a quantum hidden state fed back autoregressively}; the novelty claim in \S8.4 rests on that combination - with \citet{huang2026fully} and \citet{balauca2025practical} as the two nearest neighbours - and is stated narrowly.

\textbf{Dissipative and open quantum neural networks.} Dissipation as a computational resource has been proposed for state preparation and computation \citep{verstraete2009quantum,kraus2008preparation} and for layered quantum neural networks whose layers are completely positive maps \citep{beer2020training}. Our use is specific: a learned, outcome-unconditioned channel inside a recurrent loop, whose role is to make forgetting learnable.

\textbf{Classical analogues.} Unitary recurrent networks \citep{arjovsky2016unitary,wisdom2016full,jing2017tunable} were introduced to stabilize gradients; the difficulty of forgetting in norm-preserving recurrences is known in that literature. Gated recurrences \citep{cho2014learning} and input-dependent (selective) state-space models \citep{gu2022efficiently,gu2023mamba,orvieto2023resurrecting} are the classical counterparts of our token-conditioned channel. Classically, UMW-G is an input-switched Kraus-map state-space model with a quadratic readout; we use this correspondence to state the model's classical simulability plainly.

\textbf{Information bounds.} Holevo's theorem \citep{holevo1973bounds} bounds the accessible information about a classical variable encoded in n qubits by n bits; Nayak's bound \citep{nayak1999optimal} tightens this for random-access decoding, which is the structure of our recall benchmark (\S6).

\section{Models, controls, and protocol}\label{sec:models}
\subsection{The architecture (Figure~\ref{fig:1})}\label{sec:3-1}
A memory register of n qubits holds $\rho_t$. Each input byte $x_t \in \{0,\ldots,255\}$ selects a channel $\mathcal{E}_{x_t}$; the next byte is emitted by a 256-outcome POVM. Teacher-forced scoring, free-running generation, and the hardware protocol are the same computation:

\begin{figure}[t]\centering
\includegraphics[width=\linewidth]{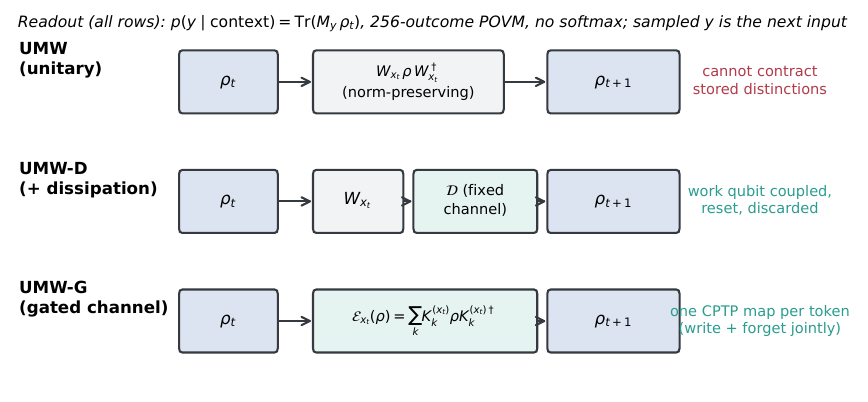}
\caption{The three memory dynamics compared. UMW: unitary conjugation (norm-preserving, cannot contract stored distinctions). UMW-D: unitary write followed by a fixed learned dissipation channel (a work qubit coupled, reset, and discarded). UMW-G: one learned CPTP map per token, writing and forgetting jointly. All rows read out with the same 256-outcome Born-rule POVM, with no softmax; the sampled byte is the next input.}
\label{fig:1}\end{figure}

\begin{align*}
  \rho_0 = |0\ldots0\rangle\langle 0\ldots0| \\
  p(y_t \mid x_{<t}) = \mathrm{Tr}(M_y\rho_t)                       &\quad\text{(Born-rule readout, no softmax)} \\
  \rho_{t+1} = \mathcal{E}_{x_t}(\rho_t)                             &\quad\text{(one channel per token)} \\
\end{align*}

Three members of the family are compared.

\begin{itemize}
  \item \textbf{UMW (unitary memory).} $\mathcal{E}_x(\rho) = W_x\rho W_x^\dagger$ with $W_x$ unitary. The memory is never measured; a separately measured work register carries the output in the recall experiments.
  \item \textbf{UMW-D (unitary write + fixed dissipation).} $\mathcal{E}_x(\rho) = \mathcal{D}(W_x\rho W_x^\dagger)$, where $\mathcal{D}(\sigma) = \mathrm{Tr}_W[V(\sigma\otimes|0\rangle\langle 0|_W)V^\dagger]$ couples the memory to a fresh work qubit through a learned unitary V and discards it. $\mathcal{D}$ is token-independent, outcome-unconditioned, and CPTP by construction.
  \item \textbf{UMW-G (gated channel).} $\mathcal{E}_x(\rho) = \sum_k K^{(x)}_k\rho K^{(x)\dagger}_k$, one learned CPTP map per token, realized as $\mathrm{Tr}_W[V_x(\rho\otimes|0\rangle\langle 0|_W)V_x^\dagger]$: writing and forgetting are learned jointly and conditioned on the input. We describe this as an input-conditioned dissipative channel playing a role analogous to a classical recurrent gate; we do not claim an exact correspondence with the GRU update equations.
\end{itemize}

The POVM is parameterized as $M_y = S G_y S$ with $G_y = R_y^\dagger R_y$ positive semidefinite and $S = (\sum_y G_y)^{-1/2}$, so that $\sum_y M_y = I$ exactly. All parameters are trained by gradient descent on the negative log-likelihood of the next byte, in exact density-matrix simulation.

\subsection{Control baselines}\label{sec:3-2}
The two controls are the backbone of every architectural claim, so we specify them exactly. Each control is the \emph{same model class and parameterization} evaluated under a different state-update rule, and each is \textbf{retrained from scratch under that rule} (so parameter \emph{values} are not shared with the recurrent model; architecture and parameter count are) - never obtained by truncating a trained recurrent model at inference.

\begin{itemize}
  \item \textbf{h1 (one-token control):} $\rho_{t+1} = \mathcal{E}_{x_t}(\rho_0)$. The memory is rebuilt from the previous token alone at every step; the model can represent at most a bigram-class predictor.
  \item \textbf{h2 (two-token control):} $\rho_{t+1} = \mathcal{E}_{x_t}(\mathcal{E}_{x_{t-1}}(\rho_0))$. The memory is rebuilt from the previous two tokens.
  \item \textbf{memoryless:} $\rho_{t+1} = \rho_0$ (sanity control; it must recover the unigram).
\end{itemize}

Parameter counts are therefore \emph{exactly} matched: at n = 4, UMW-G and each of its controls have 655,360 trainable parameters (UMW-D has $2.5\times$ fewer, and its controls match it). All arms share the corpus, split, context length, optimizer (Adam, learning rate 0.01, batch 32), and step budget (4,000--6,000 steps); seeds are reported per arm, and all evaluation uses the full held-out test set. For UMW-D we additionally trained a \emph{channel-matched} h1 (the same dissipation channel present, reset dynamics) so that the comparison isolates recurrence rather than the presence of the channel.

\subsection{Corpus, scoring, and classical baselines}\label{sec:3-3}
The corpus is the body of \emph{Alice's Adventures in Wonderland} as raw bytes (\path{data/alice_body.bin}), split 80/10/10; training uses windows of CTX = 12 bytes (24 and 48 in \S5.3); the full test set is 15,097 windows (181,164 scored predictions). Classical baselines were recomputed in-run on the same split: add-one unigram 3.2433 nats/byte; tuned interpolated bigram 2.3406. Stronger historical baselines on the same corpus from earlier milestones are a backoff bigram at 1.8614, a trigram at 1.5887, and a 4-gram at 1.5125 nats/byte. All quantum models in this paper are well above the trigram and 4-gram.

\subsection{Pre-registration as methodology}\label{sec:3-4}
Each milestone's hypotheses, metrics, and pass/fail bars were committed to \path{docs/m*_hypotheses.md} (one file per milestone) before the corresponding runs (first commits 2026-08-29 for M18--M21, 2026-08-30 for M23--M25, 2026-08-31 for M26 and M28, 2026-09-01 for M31). Amendments were appended after a result and before the next run, dated, and are visible in the files; where a bar was changed the original is retained. Appendix~\ref{app:b} lists every registered bar in this paper with its outcome. We do not treat pre-registration as evidence in itself; its role here is to make the reported negatives and the few positives equally visible.

\section{Why closed unitary recurrence cannot learn to forget}\label{sec:unitary}
\subsection{The precise statement}\label{sec:4-1}
Let $\rho$ and $\sigma$ be two memory states that encode different contexts. Under a common unitary W, the trace distance is invariant, $\tfrac12\|W\rho W^\dagger - W\sigma W^\dagger\|_1 = \tfrac12\|\rho-\sigma\|_1$, as are the spectra and hence the von Neumann entropies of both states; the map $\rho \mapsto W\rho W^\dagger$ is injective on states. Consequently a closed unitary recurrence can \emph{transform} stored distinctions - a token-conditioned $W_x$ rotates the code subspace differently for different inputs - but it cannot \emph{contract} any pair of distinct stored states toward each other, cannot reset the memory to a fixed state independent of its contents, and cannot implement any many-to-one map. A general CPTP channel can. Three properties should be kept apart: every CPTP map is non-expansive in trace distance; a non-unitary channel can \emph{strictly} contract the distance between distinct states; and some channels have attracting fixed points, up to the extreme case of a unique fixed point to which every input converges (a complete reset). All three behaviours are available within the class $\mathrm{Tr}_W[V(\rho\otimes|0\rangle\langle 0|_W)V^\dagger]$ used here, and none is available to unitary conjugation. Whether a \emph{trained} channel has an attracting fixed point is an empirical question; \S5.1 reports the measured spectrum of the learned channel (one conserved mode, finite half-lives for the rest).

This is the whole of the "cannot forget" argument, and its qualification: the obstruction is to the \emph{channel} that forgetting requires, not to any particular readout. It implies nothing about unitary models with unbounded auxiliary registers or about architectures outside the fixed-width recurrent setting studied here, and we do not claim a general impossibility theorem for unitary quantum language models.

\subsection{Consequences observed in training}\label{sec:4-2}
Two experiments isolate the consequence. First, on a marked selective-recall task (store a marked symbol, recall it after k distractors), a trained UMW keeps its symbol-conditioned memory states at pairwise fidelity $\le$ 0.009 indefinitely - the designed never-forgets property - yet its accuracy is delay-tuned: perfect at the trained delay and falling toward chance at other delays, because distractor unitaries rotate the code subspace and the POVM points where the frame sits at the trained readout time. A delay-invariant readout of width O(1) does not exist in this setting: it would require every stored state to be fixed by every distractor unitary while the "armed" state maps onto it under the same unitary, which is impossible for a unitary acting on orthogonal states; a position counter of width O(log k) removes the conflict (\path{docs/m18_delay_invariance_obstruction.md}). Training a UMW initially failed outright (0\% of the chance-to-perfect gap at every delay); diagnostics showed healthy gradient norms with low batch signal-to-noise rather than a barren plateau \citep{mcclean2018barren}, and either a delay curriculum or an annealed auxiliary recall loss passed the pre-registered bar at delays up to 8 in 3/3 seeds.

Second, as a byte language model, \textbf{unitary memory loses to its own retrained one-token control} (Figure~\ref{fig:2}, first group): full recurrence 2.5167 vs h1 2.3705 nats/byte at n = 4, and 2.5503 vs 2.3652 at n = 6. Width, a staged context curriculum (2.5040 at n = 4; 2.5599 at n = 6), replay-mixed contexts (2.7445), and per-position POVMs with $7\times$ the readout parameters (2.5221) all fail to close the gap. The reading is the one \S4.1 predicts: the control's strategy \emph{is} a reset, which the unitary model cannot represent, so it pays for every stale token it is forced to carry. Within this architecture and task, closed unitary recurrence exhibits a structural inability to implement the adaptive forgetting that byte-level language modeling rewards.

\section{Learned dissipation and token-conditioned gating}\label{sec:dissipative}
\subsection{UMW-D: forgetting as a learned physical resource}\label{sec:5-1}
With one work qubit and a learned, outcome-unconditioned dissipation channel, the model passes the test that unitary memory failed. At n = 4, three seeds reach 2.2752 / 2.2820 / 2.3050 nats/byte against channel-matched retrained h1 controls at 2.3805 / 2.3875 / 2.3958: a margin of \textbf{+0.101 $\pm$ 0.008 nats} (mean $\pm$ s.d., n = 3 seeds; range 0.091--0.106; pre-registered bar 0.05), or +0.083 $\pm$ 0.016 against the unitary model's h1 (2.3705). Two work qubits give 2.2985; n = 6 gives 2.2686. A spectral analysis of the learned channel shows a forgetting \emph{schedule}: one conserved mode, a few slow modes (6 of 256 with half-lives above four steps), and a fast-damped remainder, with mean trace-distance contraction 0.927 per step - between the two failed extremes of collapsing the memory at every token and never forgetting.

\begin{figure}[t]\centering
\includegraphics[width=\linewidth]{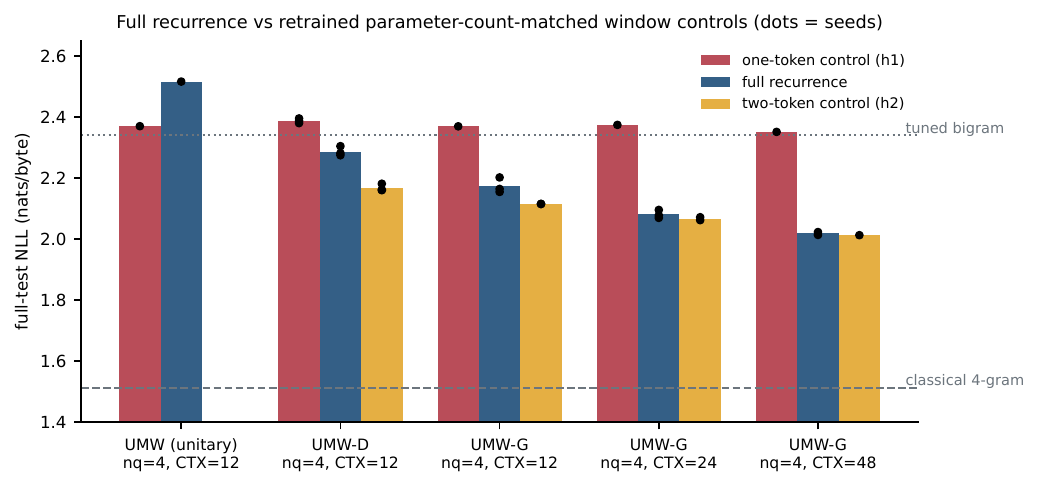}
\caption{Full recurrence against retrained one-token (h1) and two-token (h2) controls of the same parameterization and parameter count; dots are individual seeds, bars their means. Dotted and dashed lines: the tuned bigram (2.3406) and the classical 4-gram (1.5125) on the same test set.}
\label{fig:2}\end{figure}

The horizon is short, and we measured it: the retrained \textbf{two-token control beats UMW-D} (2.1606 / 2.1618 / 2.1817 vs 2.2752--2.3050), and warm-starting full training from the h2 solution degrades back to $\approx$ 2.28. Beyond one token: yes. Beyond two: not at this stage.

\subsection{UMW-G: token-conditioned dissipation}\label{sec:5-2}
Making the channel token-conditioned (one learned CPTP map per byte) improves on UMW-D by \textbf{+0.113 $\pm$ 0.025 nats} (3/3 seeds; bar 0.05) and more than doubles the beyond-one-token margin: UMW-G reaches 2.1549 / 2.1647 / 2.2024 (mean 2.1740 $\pm$ 0.025) against a retrained, parameter-count-matched h1 at 2.3701, a margin of \textbf{+0.196 $\pm$ 0.025 nats} (range 0.168--0.215). The best canonical-protocol result in the paper is 2.1500 nats/byte, obtained with the annealed entropy regularizer of \S7 (3 seeds: 2.1500 / 2.1552 / 2.1609 vs 2.1549 / 2.1647 / 2.2024 without it); the regularizer improves NLL as a side effect of the entropy objective.

The two-token control is \textbf{not beaten}: at CTX = 12 the retrained h2 reaches 2.1151 / 2.1159, and the recurrent model sits 0.058 $\pm$ 0.025 nats above it (best seed 0.039). Two attempts to add position structure failed: per-position POVMs (\S4.2) and phase-clocked channel banks (P = 2: 2.2215 / 2.2305; P = 3: 2.2965, all worse than the single bank).

\subsection{Context length: recurrence catches the two-token window and stops}\label{sec:5-3}
Controls were retrained per context length (Figure~\ref{fig:2}, right groups). At CTX = 24, full recurrence reaches 2.0697 / 2.0780 / 2.0958 against h1 2.3747 (\textbf{+0.294 $\pm$ 0.013}) and h2 2.0623 / 2.0719 (gap $-$0.014 $\pm$ 0.013); at CTX = 48, 2.0139 / 2.0233 against h1 2.3517 (\textbf{+0.333 $\pm$ 0.007}) and h2 2.0131 (gap $-$0.005 $\pm$ 0.007). Wider registers at CTX = 24 (n = 5: 2.0205 vs its h2 2.0079; n = 6: 2.0052 vs 2.0058) show the same parity. Two caveats are stated: cross-context numbers are confounded by throughput (a longer window delivers more tokens per step under a fixed step budget), so only within-context comparisons are controlled; and the number of seeds is small (2--3). Within those limits, every context length tested gives the same verdict - beyond one token, decisively; beyond two, parity - which motivates the capacity measurements of \S6.

\section{Capacity: what a small quantum memory can retrieve}\label{sec:capacity}
\subsection{Definitions}\label{sec:6-1}
We separate four quantities that the informal phrase "two-token wall" conflates.

\begin{itemize}
  \item \textbf{The encoded classical variable.} On the recall benchmark a document is \texttt{k1=v1; ... k\_K=v\_K;} followed by K queries \texttt{k?v;} (one per stored key, in random order). The generator draws the K keys \emph{without replacement} from 16 letters and the K values independently and uniformly from 10 digits. Because the queries address keys, not positions, the variable the memory must represent is the key $\to$ value map X (an unordered set of K pairs), with entropy $H(X) = \log_2\binom{16}{K} + K\log_2 10 \approx 7.3,\,13.6,\,19.1$ bits for K = 1, 2, 3. The \emph{value content} that a query must return is $K\log_2 10 \approx 3.3,\,6.6,\,10.0$ bits; the key set ($\log_2\binom{16}{K} \approx 4.0,\,6.9,\,9.1$ bits) must also be represented in some form, since the memory has to respond correctly to whichever stored key arrives.
  \item \textbf{The quantum memory state.} After the write phase the register holds $\rho_X$ on n qubits.
  \item \textbf{Accessible information.} For any single measurement, Holevo's bound gives $I(X;\mathrm{outcome}) \le S(\bar\rho) \le n$ bits. Because the query names which value to decode, the task contains a random-access code: restricted to any fixed key set it \emph{is} a quantum random-access encoding of the K values, so Nayak's bound is a necessary condition - recovering any one of m encoded bits with per-bit success probability p requires $n \ge (1 - H_2(p))\,m$ qubits. With $m = K\log_2 10$ this excludes near-perfect recall of three values at n $\le$ 9 and of two values at n $\le$ 6, but it is uninformative at the 0.5 accuracy bar, where $1 - H_2(p)$ is small.
  \item \textbf{Task success.} We score answer-byte accuracy (and NLL) at the query positions only; the value marginal fixes the floor at accuracy 0.10 (NLL ln 10 = 2.303), and an n-gram model or the h2 control sits exactly there because the key--value binding is outside their window. The pre-registered "solved" bar is accuracy $\ge$ 0.5.
\end{itemize}

We report the largest K* whose accuracy exceeds the bar at each width; it is a coarse, benchmark-specific quantity (accuracy above 0.5 does not imply full retrieval, and K moves in steps of one stored pair).

\subsection{The ladder}\label{sec:6-2}
\begin{figure}[t]\centering
\includegraphics[width=\linewidth]{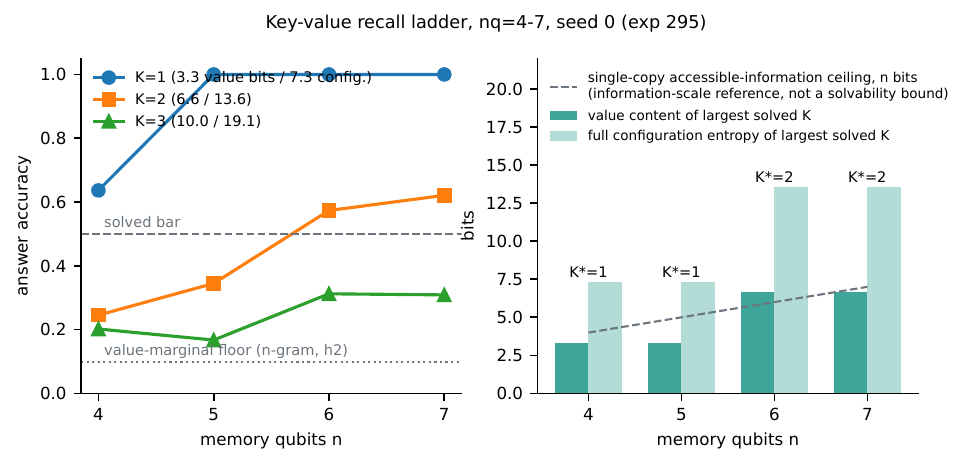}
\caption{Key--value recall ladder (exp 295, single seed). Left: answer accuracy against width for $K=1,2,3$ stored pairs, with the 0.5 bar and the value-marginal floor. Right: value content and full configuration entropy of the largest $K$ solved at each width, with the single-copy accessible-information ceiling of $n$ bits drawn as an information-scale reference only, not as a task-solvability threshold.}
\label{fig:3}\end{figure}

Table~\ref{tab:ladder} and Figure~\ref{fig:3} give the answer accuracies (exp 295; single seed, 6,000 steps, dense channels, evaluated on 1,500 held-out documents).

\begin{table}[t]\centering\small
\begin{tabular}{lcccc}
\toprule
K (value bits / configuration bits) & n = 4 & n = 5 & n = 6 & n = 7 \\
\midrule
1 (3.3 / 7.3) & \textbf{0.637} & \textbf{1.000} & \textbf{1.000} & \textbf{1.000} \\
2 (6.6 / 13.6) & 0.245 & 0.345 & \textbf{0.573} & \textbf{0.621} \\
3 (10.0 / 19.1) & 0.202 & 0.167 & 0.312 & 0.309 \\
\bottomrule
\end{tabular}
\caption{Key--value recall ladder (exp 295; answer accuracy, single seed, 6{,}000 steps, dense channels). Bold: above the pre-registered 0.5 bar.}
\label{tab:ladder}
\end{table}

Accuracy generally increases with width, though at K = 3 there are small decreases from n = 4 to 5 (0.202 to 0.167) and from n = 6 to 7 (0.312 to 0.309). By the pre-registered accuracy $\ge$ 0.5 bar, one pair is solved already at n = 4 (0.637), two pairs from n = 6 (with partial retrieval, 0.57--0.62), and three pairs are not solved through n = 7. Thus $K^*$ is 1, 1, 2, 2 across n = 4--7. Over four widths with one seed per cell and that hard threshold, these data support an \textbf{empirical, nondecreasing threshold shift in usable recall capacity} over n = 4--7 on this benchmark; they do not support a numerical rate of bits per qubit, and we make no such estimate (an earlier summary of these data as "$\approx$ 1--1.5 retrievable bits per qubit" is withdrawn). The bounds of \S6.1 are consistent with the ladder - they forbid perfect three-pair recall at n = 7 - but they do not by themselves predict the accuracies observed at the bar. The measurement is specific to this architecture, training regime, and task family. A design error is recorded: an earlier six-pair task had configuration entropy ($\approx$ 33 bits) far above the single-copy accessible-information ceiling at the tested widths (4--5 qubits) and empirically remained at the floor in every arm. Since the random-access task does not require recovering the whole configuration at once, this is not an impossibility proof for the 0.5 bar; it motivated adding a capacity sanity check to later registrations.

\subsection{Relation to the two-token horizon on text}\label{sec:6-3}
The empirical unigram entropy of the corpus is 3.24 nats = 4.7 bits per byte, so a two-byte context carries up to $\approx$ 9 bits, and the 4-gram's 1.51 nats/byte shows that substantially more than two bytes of structure is present. With 4--8 memory qubits, the accessible information per readout is at most 4--8 bits by Holevo, and the ladder shows that two sharp bindings (6.6 bits of value content; 13.6 bits of configuration) are retrieved only partially at n = 6--7. The persistent parity with the retrained two-token window across architectures and contexts is therefore \textbf{consistent with an information-capacity bottleneck of the scale suggested by the Holevo bound}. We do not claim to have established the bound as the cause: the capacity measurement is on a synthetic task, the text horizon is measured against a window control, and optimization effects on text cannot be excluded by these experiments alone. The evidence supports the interpretation; it does not prove it. The phrase "two-token wall" is used below only as shorthand for the measured parity.

\section{A stationary memory-entropy separation (secondary result)}\label{sec:entropy}
\textbf{Setup.} The perturbed coin is a two-state process whose output flips with probability p; its statistical complexity is $C_\mu = \ln 2$ exactly (every exact classical predictor needs one bit of memory), and the minimal quantum memory entropy $C_q(p)$ is known in closed form (0.173, 0.057, 0.018 nats at p = 0.3, 0.4, 0.45). One- and two-qubit UMW-G models were trained on raw streams. A \emph{floor gate} requires each model to predict at the process's entropy rate (excess NLL $\le$ 0.002 nats): 18/18 runs passed without the regularizer and 27/27 with it.

\begin{figure}[t]\centering
\includegraphics[width=0.8\linewidth]{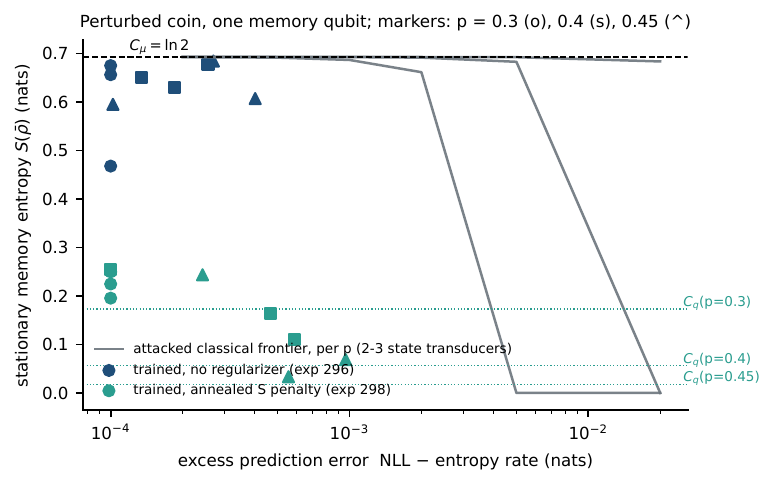}
\caption{Perturbed coin, one memory qubit: stationary memory entropy $S(\bar\rho)$ against excess prediction error, for trained models without (dark) and with (teal) the annealed entropy penalty; markers denote $p = 0.3$ (circle), $0.4$ (square), $0.45$ (triangle). Grey curves: the attacked classical frontier per $p$; dashed: $C_\mu = \ln 2$; dotted: the quantum optima $C_q(p)$.}
\label{fig:4}\end{figure}

\textbf{Classical attack.} The comparison class is classical predictors at matched prediction error, found by a Lagrangian optimization (multi-restart Nelder--Mead) over 2- and 3-state classical transducers with exact pair-chain evaluation of both their prediction error $\varepsilon$ and their stationary memory entropy. The resulting entropy--error frontier is flat at $C_\mu$ for small $\varepsilon$ (0.6931 at $\varepsilon \le 0.002$ for p = 0.3; $\ge$ 0.6918 at every trained model's measured $\varepsilon$) and collapses toward zero only at $\varepsilon \gtrsim 0.005$--$0.02$, where a memoryless predictor suffices. All trained models sit at $\varepsilon \le 0.002$, in the flat regime.

\textbf{Results (Figure~\ref{fig:4}).} Without any entropy objective, all nine one-qubit seeds land below the attacked frontier at their own $\varepsilon$: margins 0.008--0.225 nats (mean 0.066 $\pm$ 0.067, median 0.041; four of nine exceed 0.05). With an annealed penalty $\lambda_t\,S(\bar\rho)$ that is driven to zero before training ends (so the final objective is pure prediction), the margins become 0.438--0.658 nats (mean 0.521 $\pm$ 0.083, \textbf{9/9 seeds}), with the best run at S = 0.0335 nats against $C_q$ = 0.0175 at p = 0.45. Three pre-registered bars failed and are reported: seed-reliable emergent margins above 0.05 (4/9); reaching $C_q$ + 0.1 in 3/3 seeds in at least two settings (3/3, 2/3, 1/3 across p); and compression of a two-qubit register below $C_\mu$ (entropy spreads \emph{above} $C_\mu$ when spare width is available).

\textbf{Scope.} This is a \emph{stationary quantum-memory entropy separation from the tested classical predictive models at matched prediction error}, anchored at zero error by the exact theorem $C_\mu = \ln 2$. It is not a speed, sample-complexity, parameter, or accuracy advantage; the thermodynamic reading is only that $S(\bar\rho)$ is the quantity that bounds the minimal work of erasure for the simulator's memory, and no energy or work measurement was performed; and the classical search is heuristic beyond three-state transducers. The result is included because it shows that gradient training reaches the regime the analytic constructions predicted; it is not the main claim of the paper.

\section{Hardware validation (IBM Heron, \texttt{ibm\_marrakesh})}\label{sec:hardware}
\subsection{Compilation and local validation}\label{sec:8-1}
The dense channels were compiled to hardware-shaped circuits by task-aware distillation (\S9): each byte's channel became a depth-12 circuit of single-qubit Ry/Rz rotations and a line of CZ gates on 4 memory + 1 work qubits (120 parameters per channel against 2,048 for the dense map). Before any submission, every per-token unitary built in Qiskit was checked against the simulator's construction to a maximum deviation of $5\times10^{-10}$, and Aer sampling of every circuit was checked against the simulator's predicted distribution (total variation distance 0.013 $\pm$ 0.003 at 8,192 shots, consistent with sampling error). This gate caught one bit-ordering bug before quantum time was spent. All runs used 4,096 shots per circuit; the total charge across Jobs A--C was $\approx$ 150 s. Job identifiers: A \texttt{daardrmrrl7c7385vf10}, A$'$ \texttt{daash9j4clkc73fi44ag}, B \texttt{daatbjbvpcac73dd029g}, B2 \texttt{daatg51l216s739oo84g}, B3 \texttt{daatl9pl216s739oohh0}.

\begin{figure}[t]\centering
\includegraphics[width=\linewidth]{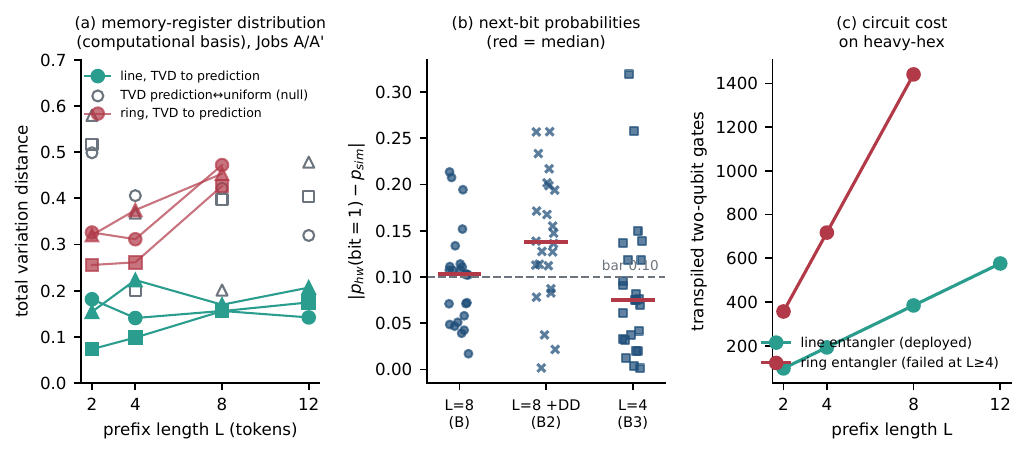}
\caption{Hardware validation on \texttt{ibm\_marrakesh}. (a) Total variation distance between the measured computational-basis memory-register distribution and the simulator's predicted diagonal, for line (Job A$'$) and ring (Job A) circuits, against the distance between the prediction and the uniform null (hollow markers). (b) Per-bit $|p_{\mathrm{hw}}(\mathrm{bit}=1) - p_{\mathrm{sim}}|$ for Jobs B (L=8), B2 (L=8 with dynamical decoupling and twirling) and B3 (L=4); red bars are medians, dashed line the 0.10 bar. (c) Transpiled two-qubit gate counts for the two entanglers.}
\label{fig:5}\end{figure}

Three levels of validation are distinguished (Figure~\ref{fig:5}). At every level, what ran on the processor was the full quantum state evolution (channel plus work-qubit reset per token) and the measurement; what remained classical was circuit construction, shot counting, and - for generation - the selection of the next circuit and the sampling of a bit from a measured probability.

\subsection{Level 1: memory-register distribution validation (Jobs A, A$'$)}\label{sec:8-2}
Each circuit replays a prefix of L tokens (channel, then reset of the work qubit, per token) and measures the memory register. The test statistic is the total variation distance between the measured 16-bin distribution and the simulator's predicted memory diagonal, compared with the distance between the prediction and the uniform distribution (the noise null); a circuit "shows signal" when the former is smaller.

The first attempt (Job A) used ring-entangler circuits; on the heavy-hex topology these transpiled to 354--1,452 two-qubit gates ($\approx 3\times$ the design count) and showed signal in only 4 of 9 circuits, passing the pre-registered bar at L = 2 only - a failure recorded and diagnosed. Re-distilling with a line entangler (identical simulated quality, 2.2746 vs 2.2735 nats/byte; $3.7\times$ fewer transpiled gates, 96--576) gave \textbf{signal in 12/12 circuits} across three prefixes and L $\in$ {2, 4, 8, 12}: TVD to the prediction 0.156 $\pm$ 0.041 (range 0.073--0.223; by length 0.136, 0.154, 0.161, 0.175 at L = 2, 4, 8, 12) against prediction-to-uniform distances of 0.20--0.58. The measured memory-register distribution tracks the simulator's prediction through the full 12-token context, and the computational-basis statistics remain distinguishable from the uniform null at every length. No tomography was performed: nothing is claimed about off-diagonal elements or the coherence of the hardware state.

\subsection{Level 2: next-bit prediction validation (Jobs B, B2, B3)}\label{sec:8-3}
Next-byte probabilities were measured through a distilled binary-tree readout: eight stages, each a 5-qubit circuit (memory + one ancilla) selected by the bits already fixed, so that $p(y) = \prod_s p(\mathrm{bit}_s \mid \mathrm{bits}_{<s})$. This tree is a distilled hardware approximation to the model's 256-outcome vocabulary POVM, implemented as a product of eight physically measured conditional bit probabilities: the mathematical model is defined with the 256-outcome POVM, the hardware implements the distilled binary tree, each conditional bit probability is physically measured, their product defines the hardware token probability, and no learned classical vocabulary projection or softmax is used. The processor never performs a single monolithic 256-outcome measurement, and the distillation has a measurable approximation error (simulated NLL 2.3604 for the tree vs 2.2735 for the dense POVM on the same channels). The tree was trained in simulation against the compiled channels (25,500 angles at depth 10; simulated NLL 2.3604 vs the student's 2.2735). Because each stage runs on a fresh copy of the prefix replay, the protocol is back-action-free by construction. Jointly fine-tuning channels and tree readout in simulation lowers the tree model from 2.3604 to 2.2018 nats/byte (exp 308; registered bar +0.10 against the staged tree, passed); the hardware jobs reported in this section used the staged (pre-fine-tuning) parameters. The test compares the measured p(bit = 1) with the simulator's for 24 bit decisions (3 teacher-forced cases $\times$ 8 stages); the registered bars were a median $|\Delta p|$ $\le$ 0.10 and $\ge$ 20/24 agreeing decisions.

At 8-token prefixes (Job B): median $|\Delta p|$ = 0.1028 (bar missed by 0.003; \textbf{fail}) with 21/24 decisions agreeing (\textbf{pass}). Adding dynamical decoupling and Pauli twirling (Job B2) made the per-bit error \emph{worse} (median 0.1374, 21/24) - a recorded negative. At 4-token prefixes (Job B3): median $|\Delta p|$ = 0.0749 and 22/24 decisions - \textbf{both bars pass}, with the per-bit error decreasing (median 0.103 $\to$ 0.075) as the two-qubit gate count halves.

\subsection{Level 3: autoregressive generation (Job C)}\label{sec:8-4}
Generation ran hardware-in-the-loop with the simulator's own generation semantics (Figure~\ref{fig:6}): for each of the eight bits of a byte, a fresh circuit (prefix replay plus the current tree-stage readout) was executed, the bit was sampled classically from the measured probability, and the sampled bit selected the next stage's circuit; after eight bits the byte was appended to the prompt and fed back as the next input. Every state transition and every bit probability was physically executed; the classical parts were shot counting, bit sampling, circuit selection, and the prompt buffer. From the prompt \texttt{t wa} the processor generated \texttt{r}, \texttt{0x14}, \texttt{l} (24 circuits; archived with per-stage probabilities in \path{hardware_candidate_m30/job_c_generation.json}). A replication script, \path{hardware_candidate_m30/job_c2.py}, performs the same generation with full per-stage archiving (job identifiers, raw counts, measured and simulated probabilities, sampled bits, circuit metadata).

\begin{figure}[t]\centering
\includegraphics[width=\linewidth]{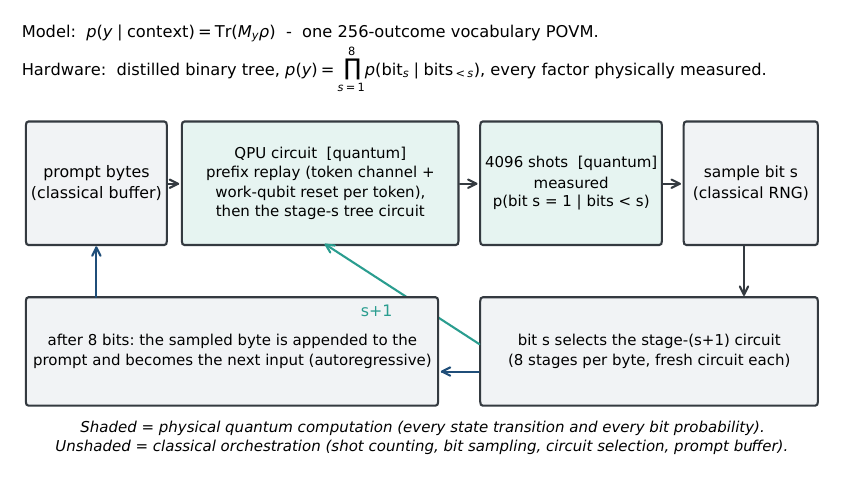}
\caption{Hardware-in-the-loop generation. The model is defined with a single 256-outcome vocabulary POVM; the hardware implements a distilled binary-tree approximation in which each of eight conditional bit probabilities is physically measured and their product defines the token probability. Shaded boxes are physical quantum computation (every state transition and every bit probability); unshaded boxes are classical orchestration (shot counting, bit sampling, circuit selection, the prompt buffer). No learned classical vocabulary projection or softmax is used.}
\label{fig:6}\end{figure}

The generated string is letter-shaped and weak, which is what the model's NLL class plus device noise predicts; it is \textbf{not} evidence of language quality. The point of Level 3 is end-to-end physical execution semantics. The novelty claim is correspondingly narrow. Persistent quantum recurrent state on NISQ hardware exists in prior work \citep[unitary recurrence, forecasting outputs]{huang2026fully}, and generative quantum language modeling has been run on IBM hardware with a classical vocabulary projection \citep{balauca2025practical}. To our knowledge, no prior work combines, on physical hardware, a trained recurrent model for language whose persistent hidden state is quantum and whose token-conditioned dynamics are learned dissipative channels, with token probabilities constructed entirely from physically measured Born-rule probabilities through the distilled quantum readout - no learned classical vocabulary projection or softmax - and the sampled token fed back autoregressively. The claim is to that combination, not to any of its components.

\subsection{Noise margin and hardware-native training}\label{sec:8-5}
A noise-gate simulation of the compiled model (two-qubit depolarizing error $e$ and readout error $r$) gives byte NLL 2.29 at $(e, r) = (0.005, 0.01)$, 2.34 at (0.02, 0.03), 2.38--2.46 at $e = 0.05$, and 2.47--2.55 at $e = 0.1$, against the unigram 3.24 - roughly 0.7--0.9 nats of margin at present device error rates. Separately, simultaneous-perturbation stochastic approximation over shot-sampled objectives (no gradients) trains the 80-parameter one-qubit model to its floor (3/3 seeds, 0.6741--0.6743 vs 0.6730; $\approx$ $0.7\times10^{9}$ shots) but leaves the 655k-parameter dense model at random (5.51 $\approx$ ln 256 after $2.1\times10^{9}$ shots) - the standard variational-training reality, measured for this architecture - while holding the compiled 30,720-angle model within +0.003 nats over $1.66\times10^{6}$ circuit executions (2/2 seeds; bar +0.02). The hardware pipeline that follows is: pretrain in simulation $\to$ task-aware compilation $\to$ hardware fine-tuning. No task-level speedup is claimed.

\section{Training beyond exact density-matrix simulation, and what width and depth buy}\label{sec:scaling}
Tracking the purification factor A ($\rho = AA^\dagger$, rank $\le 2^n$) instead of $\rho$, and applying gates to A directly, is exact at full rank (equivalence to $\rho$-based training at $10^{-10}$; the registered correctness gate at n = 4 passed at 2.3114 nats/byte) and enables controlled low-rank training at 7--8 qubits under a rank cap, with discarded weight below $10^{-6}$ in the reported runs. Appendix~\ref{app:a} describes the representation and the numerical measures it required. Here we report what it enabled, all negatives.

\begin{figure}[t]\centering
\setcounter{figure}{0}\renewcommand{\thefigure}{S\arabic{figure}}
\includegraphics[width=\linewidth]{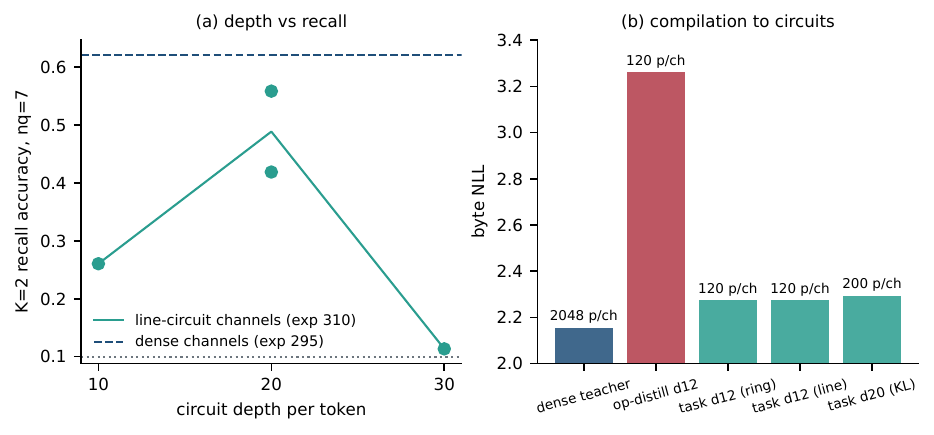}
\caption{Supplementary. (a) $K=2$ recall accuracy at $n=7$ against circuit depth per token for line-circuit channels (exp 310; dots are seeds) and the dense-channel reference (exp 295); the depth-30 point is from the follow-up campaign and is discussed in the footnote to \S\ref{sec:scaling}. Points are not interpolated. (b) Byte NLL of the dense teacher, an operator-level distilled student, and task-aware students at depths 12 and 20, with parameters per channel.}
\label{fig:S1}\end{figure}

\begin{itemize}
  \item \textbf{Byte NLL does not improve with width under feasible budgets.} At CTX = 12, n = 7 reaches 2.2520 and n = 8 2.2985 nats/byte with circuit-native channels, against 2.1549--2.1500 at n = 4; the registered CTX = 24 test at n = 7 gave 2.2726, \emph{worse} than the same model at CTX = 12. Together with \S5.3, this is the third scaling negative: width, context, and (below) depth do not buy performance on byte-level English at these scales, consistent with the corpus's two-byte dominance and the capacity picture of \S6.
  \item \textbf{Compilation costs $\approx$ 0.12--0.15 nats.} Operator-level distillation of the dense channels into shallow circuits fails (fidelities 0.25--0.44 at depths 4--12; NLL 3.26--3.37, unigram-class), while task-aware distillation succeeds at 10--17$\times$ compression (depth-12 students 2.2735 ring / 2.2746 line; depth-20 KL students 2.2929 / 2.2940; one depth-12 KL student missed the bar by 0.001). The model needs its channels' action on visited states, not the full operators.
  \item \textbf{Recall capacity depends on circuit depth where text does not.} K = 2 recall at n = 7 with line circuits rises from 0.260 at depth 10 to 0.489 $\pm$ 0.099 at depth 20 (two seeds: 0.558, 0.419), still below the dense channels' 0.621, while byte NLL is nearly depth-insensitive. We summarize this as \emph{statistical structure is depth-cheap; sharp memory is depth-hungry} - and on hardware depth is coherence.\footnotemark{} The circuit-native channels used throughout \S8 are therefore the cheapest adequate representation for text and the binding constraint for memory.
\end{itemize}

\footnotetext{A depth-30 run in the follow-up campaign did not continue the trend (0.113, chance), and its diagnosis - an optimization effect, not an expressivity limit - is reported separately. The depth observation here is confined to the 10 $\to$ 20 step and two seeds.}

\section{Limitations}\label{sec:limitations}
\begin{itemize}
  \item \textbf{Corpus and task.} All language-modeling claims rest on one English corpus at byte level; the capacity ladder uses one synthetic key--value family with a single seed per cell and a coarse task-size resolution (one stored pair per step); the entropy result uses one two-state process.
  \item \textbf{Width.} 1--8 qubits. Measured scalings over so narrow a range extrapolate poorly, and our architecture-specific capacity numbers should not be read as properties of quantum memory in general.
  \item \textbf{Controls and seeds.} Controls are retrained from scratch with the same parameterization and parameter count, but seeds are few (2--3 per arm; 1 on the ladder); headline margins are reported as mean $\pm$ s.d. where more than one seed exists.
  \item \textbf{Classical attack.} Beyond three-state transducers the entropy frontier is heuristic; the anchor is exact only at zero error.
  \item \textbf{Hardware.} One device family (IBM Heron, one processor for the main jobs), 4,096 shots per circuit, $\approx$ 150 s of total charge, and compilation-specific effects (the ring-to-line change alone decided Level 1). The 4-token prediction pass and the 8-token near-miss are at the edge of the device's coherent depth for these circuits.
  \item \textbf{No computational speedup; classical simulability.} All models are exactly simulable at these widths, all training was simulated, and the hardware runs validate the stack rather than any advantage.
  \item \textbf{Language quality.} No model here approaches the classical 4-gram (1.51 nats/byte); the generated hardware text is weak.
  \item \textbf{Causality of the horizon.} The capacity interpretation of the two-token parity is consistent with the evidence, not established by it.
\end{itemize}

\section{Conclusion}\label{sec:conclusion}
Within a fixed-width recurrent architecture with Born-rule readout, a purely unitary quantum memory cannot contract or reset what it stores and, in our experiments, cannot even match a retrained one-token control; learned dissipation restores forgetting as a trainable resource and token-conditioned dissipation more than doubles the value of memory beyond one token, against architecture- and parameter-count-matched controls. That memory is sharply capacity-limited: a retrained two-token window is never beaten at 4--8 qubits, the largest solvable recall task shifts from one stored pair to two as width grows while three pairs remain out of reach at 7 qubits, and these facts are consistent with - though not proven to be caused by - accessible-information bounds. The same model, compiled to line circuits, reproduces its predicted computational-basis memory distribution on a 156-qubit processor through a 12-token context (within the pre-registered signal criterion), reproduces its next-bit probabilities within pre-registered bars at 4-token prefixes, and generates bytes with every bit drawn from a measured probability.

These models do not beat strong classical language models and are not close to doing so; the contribution is architectural, physical, and methodological. The measured threshold shift suggests that substantially larger logical quantum memories - and, by the depth--coherence trade of \S9, error-corrected ones - would be required before this architecture could represent language contexts competitive with classical baselines; we do not derive a qubit count from the present data. What the work establishes is which dynamics a trainable quantum-native recurrent model requires, how to measure what its memory can hold, and that the whole stack already runs on hardware.

\section*{Reproducibility}\label{sec:repro}\addcontentsline{toc}{section}{Reproducibility}
The supporting code, experiment records, and research materials are maintained in a private repository (tag \texttt{paper-submission-package}), which is not publicly accessible at the time of this preprint. The repository contains milestone reports M18--M31 (\path{MILESTONE_*_REPORT*.md}), pre-registrations with dated amendments (\path{docs/m*_hypotheses.md}), every experiment as a standalone script (\path{experiments/<id>_<name>/run.py}), raw result JSON and saved parameter files (\path{results/milestone*/raw/}), hardware manifests, submission records with job identifiers, per-circuit results, and the local validation scripts (\path{hardware_candidate_m30/}), the research log, and the regression suite (\path{tests/}, 77 passing with \texttt{pytest}; \path{tests/test_core.py} is the mathematical consistency gate). The environment used is pinned in \path{requirements.txt} (Python 3.13.9, numpy 2.5.1, torch 2.13.0, qiskit 2.5.1, qiskit-aer 0.17.2, qiskit-ibm-runtime 0.48.0). Seeds are the integer seeds in each result filename. The main tables and figures are reproduced by: Figure~\ref{fig:2} from \path{experiments/290_umwlm_byte}, \texttt{292\_umwg\_byte}, \texttt{293\_umwg\_position}; Table~\ref{tab:ladder} and Figure~\ref{fig:3} from \path{experiments/295_umwgk_width}; Figure~\ref{fig:4} from \path{experiments/296_entropy_advantage}, \texttt{297\_entropy\_classical\_attack}, \texttt{298\_entropy\_regularized}; Figure~\ref{fig:5} from \path{hardware_candidate_m30/job_a*.py} and \path{hardware_candidate_m30/job_b.py}; Figure~\ref{fig:6} from \path{hardware_candidate_m30/job_c.py} (replication script \path{hardware_candidate_m30/job_c2.py}); Figure~\ref{fig:S1} from \path{experiments/301_channel_distillation}, \texttt{302\_task\_distillation}, \texttt{306\_line\_distillation}, \texttt{310\_factor\_kv}; and all figures by \path{python paper/make_figures.py}. Backend calibration was not snapshotted for Jobs A--C; the transpiled two-qubit gate counts and depths per circuit are stored in each job's metadata file.

\bibliographystyle{plainnat}
\bibliography{references}

\appendix
\section{The purification-factor representation and its numerical measures}\label{app:a}
Writing $\rho = AA^\dagger$ with $A$ of shape $(d, r)$, $r \le d = 2^n$, each gate of a token circuit is applied to the rows of A in $O(d\,r)$ without forming a $d\times d$ operator; the dissipative step lifts A with the work qubit in $|0\rangle$, applies the circuit, splits the rows by work-qubit outcome into two Kraus branches $[K_0A,\,K_1A]$, and compresses the result back to rank $\le d$. At full rank the compression is an exact QR factorization (the column space has rank $\le d$), equivalent to $\rho$-based training to $10^{-10}$. With a rank cap $r_{\max} < d$ (used at n $\ge$ 7 for cost), compression keeps the top $r_{\max}$ singular directions; the discarded weight is logged at every step and stayed below $10^{-6}$ in the reported runs (validity bound $10^{-3}$). The 7--8-qubit runs of \S9 used the rank cap and are therefore controlled low-rank approximations of density-matrix training, not exact density-matrix training.

Five numerical failure modes were encountered and resolved before the reported runs: single-precision complex SVD collapse (resolved by double-precision decompositions); LAPACK non-convergence on exactly degenerate memory states, which converged models produce; an SVD hang on a pathological input (replaced by an eigendecomposition of the Gram matrix); the phase ambiguity of complex eigenvector gradients in autograd; and a jitter that broke the exactness test at full rank (resolved by separating the exact-QR and capped paths). The final scheme is a straight-through truncation: the compression basis is computed without gradients, gradients flow only through the factor, and the forward pass is exact up to the logged discard. No decomposition appears in the autograd graph. Readout probabilities are computed as Tr($M_y$ $\rho$) with $\rho$ = $AA^\dagger$ formed explicitly, and each token step is wrapped in activation checkpointing; both changes are numerically identical to the original implementation and are recorded with their measured costs in the research log.

\section{Pre-registered bars and outcomes}\label{app:b}
\footnotesize
\begin{longtable}{P{0.30\linewidth} P{0.22\linewidth} P{0.40\linewidth}}
\toprule
Milestone / hypothesis & Bar & Outcome \\
\midrule
\endhead
M18 H: UMW trainable with signal shaping & pass recall bar at delays $\le$ 8, 3/3 seeds & \textbf{pass} (curriculum and auxiliary loss) \\
M19 H1: pipeline gate & memoryless controls recover unigram & pass \\
M19 H2--H7: unitary memory beats h1 (5 variants, 2 widths) & NLL < h1 & \textbf{fail}, all variants \\
M19 H8: UMW-D beats channel-matched h1 & $\ge$ +0.05 nats & \textbf{pass} 4/4 arms (+0.091\ldots{}+0.106) \\
M19 H9: UMW-D beats h2 & NLL < h2 & \textbf{fail} (h2 wins) \\
M20 H1: UMW-G beats UMW-D & $\ge$ +0.05, 3/3 seeds & \textbf{pass} (+0.085\ldots{}+0.132) \\
M20 H2: UMW-G beats own h2 & NLL < h2 & \textbf{fail} (gap 0.039--0.087) \\
M21 H1: phase-clocked channels & NLL < single bank & \textbf{fail} \\
M21 H2/H2x: recurrence beats h2 at CTX 24/48 & NLL < h2 & \textbf{fail} (parity) \\
M23 H: K = 1 recall beats h2 by $\ge$ 0.5 nats, acc $\ge$ 0.5 & as stated & \textbf{pass} (+1.0\ldots{}+1.4 nats); K = 6 design error recorded \\
M24 H1: Kronecker channels within fidelity gate & $\le$ 0.05 nats cost & \textbf{fail} (0.12 nats) \\
M24 H2: ladder monotone in width, thresholds scale & as stated & \textbf{pass for threshold} ($K^* = 1,1,2,2$; per-$K$ accuracy not strictly monotone) \\
M24 H3: byte gains from width (Kron) & NLL < h2 at n = 7 & \textbf{fail} (attributed to H1) \\
M25 H1: emergent S < $C_\mu$ $-$ 0.05, 3/3 per setting & as stated & \textbf{fail} (4/9) \\
M25 H2: emergent S within 0.1 of $C_q$ & as stated & \textbf{fail} \\
M26 floor gate & excess NLL $\le$ 0.002 after annealing & pass 27/27 \\
M26 H1: designed S $\le$ $C_q$ + 0.1, 3/3 seeds in $\ge$ 2 settings & as stated & \textbf{fail} (3/3, 2/3, 1/3) \\
M26 H2: n = 2 register below $C_\mu$ & as stated & \textbf{fail} \\
M28 H1: SPSA trains the 1-qubit model to floor & excess $\le$ 0.01, 3/3 & \textbf{pass} \\
M28 H2: SPSA trains the dense byte model & NLL < unigram & \textbf{fail} \\
M29 gate: operator-level distillation & fidelity $\ge$ 0.99 & \textbf{fail} (0.25--0.44) \\
M29 H: task-aware compilation & $\le$ dense + 0.15 nats & \textbf{pass} 3/5 arms (one miss by 0.001) \\
M29 addendum: SPSA holds the compiled model & $\le$ +0.02 nats & \textbf{pass} 2/2 \\
M30 Job A (ring): memory signal at L = 2, 4 & signal in both & \textbf{fail} at L = 4 \\
M30 Job A$'$ (line): memory signal & signal at every L & \textbf{pass} 12/12 \\
M30 Job B (L = 8): median $|\Delta p|$ $\le$ 0.10; $\ge$ 20/24 decisions & as stated & \textbf{fail} (0.1028) / \textbf{pass} (21/24) \\
M30 Job B2 (L = 8, DD + twirling) & as B & \textbf{fail} (0.1374) / pass (21/24); worse than B \\
M30 Job B3 (L = 4) & as B & \textbf{pass} / \textbf{pass} (0.0749; 22/24) \\
M30 tree readout distillation & $\le$ student + 0.10 nats & \textbf{pass} (depth 10) \\
M31 H1: factor training exact at full rank; feasible at 7--8 qubits under a rank cap & equivalence $10^{-10}$; gate at n = 4; discard < $10^{-3}$ & \textbf{pass} \\
M31 H2: byte NLL improves at width (CTX = 24) & NLL < n = 4 record & \textbf{fail} (2.2726) \\
M31 H3: K = 2 recall $\ge$ 0.9 at n = 7 with circuits & as stated & \textbf{fail} (0.26 $\to$ 0.49 $\pm$ 0.10 with depth) \\
\bottomrule
\end{longtable}
\normalsize

\clearpage
\section{Evidence level of the main claims}\label{app:c}
\small
\begin{longtable}{P{0.30\linewidth} P{0.22\linewidth} P{0.40\linewidth}}
\toprule
Claim & Evidence level & Where \\
\midrule
\endhead
Unitary channels preserve trace distance and cannot implement many-to-one maps & mathematical (elementary) & \S4.1 \\
No O(1)-width delay-invariant unitary recall solution in the marked-recall setting & mathematical (short argument) & \S4.2, \path{docs/m18_delay_invariance_obstruction.md} \\
Unitary memory loses to retrained h1 in this architecture & experimental (5 variants, 2 widths, single seeds) & \S4.2 \\
UMW-D and UMW-G beat parameter-count-matched h1 & experimental (3 seeds each; mean $\pm$ s.d.) & \S5 \\
Recurrence never beats retrained h2 at 4--8 qubits & experimental (2--3 seeds per context) & \S5.3 \\
Solvable recall task shifts from one to two pairs with width; three pairs unsolved at n = 7 & experimental (single seed per cell) & \S6.2 \\
Two-token horizon is a capacity effect & empirically suggested (consistency with bounds) & \S6.3 \\
Memory-entropy separation at matched error & experimental vs heuristic attack; exact anchor at $\varepsilon = 0$ & \S7 \\
Hardware memory-register distribution (computational basis) tracks prediction through 12 tokens & experimental (12/12 circuits, pre-registered statistic; no tomography) & \S8.2 \\
Hardware next-bit probabilities within bars at L = 4 & experimental (24 decisions) & \S8.3 \\
Autoregressive hardware generation with measured probabilities & experimental (execution semantics; one archived run; replication script provided) & \S8.4 \\
Novelty of the hardware-generation combination & literature-based, stated narrowly against \citet{huang2026fully} and \citet{balauca2025practical} & \S8.4, \S2 \\
Larger logical memories needed for competitive LM & speculative / extrapolated & \S11 \\
\bottomrule
\end{longtable}
\normalsize

\end{document}